\documentclass[sigconf, anonymous=false]{acmart}
\renewcommand\footnotetextcopyrightpermission[1]{} 
\usepackage{tabularx}
\AtBeginDocument{%
  \providecommand\BibTeX{{%
    \normalfont B\kern-0.5em{\scshape i\kern-0.25em b}\kern-0.8em\TeX}}}
\begin{document}

\title{Exploring the Scope and Potential of Local Newspaper-based Dengue Surveillance in Bangladesh}


\author{Nazia Tasnim}
\authornote{Both authors contributed equally to this research.}

\affiliation{%
  \institution{Shahjalal Univ of Science and Tech}
  \state{Sylhet}
  \country{Bangladesh}
}

\email{embers1010@gmail.com}

\author{Md. Istiak Hossain Shihab}
\authornotemark[1]

\affiliation{%
  \institution{Shahjalal Univ of Science and Tech}
  \state{Sylhet}
  \country{Bangladesh}
}

\email{istiak@protonmail.ch}

\author{Moqsadur Rahman}

\affiliation{%
  \institution{Shahjalal Univ of Science and Tech}
  \state{Sylhet}
  \country{Bangladesh}
}

\email{moqsad-cse@sust.edu}

\author{Sheikh Rabiul Islam}    
\affiliation{%
  \institution{University of Hartford}
  \city{West Hartford}
  \state{Connecticut}
  \country{USA}}
\email{shislam@hartford.edu}

\author{Mohammad Ruhul Amin}    
\affiliation{%
  \institution{Fordham University}
  \city{New York City}
  \state{New York}
  \country{USA}}
\email{mamin17@fordham.edu}

\renewcommand{\shortauthors}{Nazia and Istiak, et al.}
\begin{abstract}

Dengue fever has been considered to be one of the global public health problems of the twenty-first century, especially in tropical and subtropical countries of the global south. The high morbidity and mortality rates of Dengue fever impose a huge economic and health burden for middle and low-income countries. 
It is so prevalent in such regions that enforcing a granular level of surveillance is quite impossible. Therefore, it is crucial to explore an alternative cost-effective solution that can provide updates of the ongoing situation in a timely manner. In this paper, we explore the scope and potential of a local newspaper-based dengue surveillance system, using well-known data-mining techniques, in Bangladesh from the analysis of the news contents written in the native language. In addition,
we explain the working procedure of developing a novel database, using human-in-the-loop technique, for further analysis, and classification of dengue and its intervention-related news. Our classification method has an f-score of 91.45\%, and matches the ground truth of reported cases quite closely.
Based on the dengue and intervention-related news, we identified the regions where more intervention efforts are needed to reduce the rate of dengue infection.
A demo of this project can be accessed at: http://erdos.dsm.fordham.edu:3009/
\end{abstract}



\maketitle

\section{Background and Related Studies}

Dengue fever is a common arboviral infection~\cite{whoreport2000} of the Flaviviridae family~\cite{simmons2012dengue} which can be caused by four antigenically different dengue virus serotypes. It is spread to people mostly through two different vectors, primarily found in tropical and subtropical regions~\cite{rigau1998dengue}. It can induce a range of clinical events that may result in Dengue Fever (DF) or more severe variants such as Dengue Hemorrhagic Fever (DHF) and Dengue Shock Syndrome (DSS) \cite{world2009dengue}. We discuss the existing surveillance systems in the \textbf{Table \ref{tab:existingSurveillanceSystem}}

Dengue in Bangladesh is a seasonal disease which, since its primary breakout in 2000 with 1.7\% fatality in three of the main cities, has become a serious health problem owing to the country's dense population, socioeconomic demography, and environmental characteristics ~\cite{yunus2001dengue}. 
Currently, passive surveillance, on which the country heavily depends on, is insufficient to combat dengue fever in a geographical area like Bangladesh ~\cite{low2015dengue}. Moreover, the disparity in the distribution of hospitals and medical facilities around the country, as well as the tendency among lower-income people to seek medical treatment only when their health condition is life-threatening, skewed the passive surveillance data towards a specific populations~\cite{sharmin2015interaction}. 
As a result, in recent years, we believe that there has been a significant level of under-reporting from the hospital-based surveillance resulting in inefficient dengue preventive measures by the government. 

\begin{table}[t]
  \caption{Existing Surveillance Systems}
  \label{tab:existingSurveillanceSystem}
  \begin{tabular}{|p{2cm}|p{3cm}|p{2.5cm}|}
    \hline
    System Name  & Data Source & Applicability in Bangladesh \\
    \hline
    VBORNET~\cite{braks2011towards} & Multilingual disease search keyword, arthropod surveillance and information from entomologists. & Expensive program managed by ECDC which does not operate in Bangladesh. \\ \hline
    HealthMap~\cite{freifeld2008healthmap} & Online news reports, RSS feeds, and validated official alerts. & Not good with granular level of surveillance. \\ \hline
    Google Flu Trends\cite{ginsberg2009detecting} & English search queries (used for the detection of influenza) & Only works with English language. \\ \hline
    SourceSeer~\cite{rekatsinas2015sourceseer} & Newspaper articles, twitter data and RSS feed (used for forecasting outbreaks) & Depends on internet accessibility and works better for rare diseases. \\ \hline
  \end{tabular}
\end{table}

In Bangladesh, the Directorate General of Health Services (DGHS) control room gathers aggregated dengue data of hospital-compliant patients in the capital city Dhaka and generates a day-to-day report. Additionally,  Institute of Epidemiology Disease Control And Research (IEDCR), Health Population and Nutrition Sector Program (HPNSP), Access to Information (a2i) and the City Corporations collaborate to perform annual vector surveys and  take certain preventive measures~\cite{mutsuddy2019dengue} . 
However, such preventive efforts are limited by expensive out-of-pocket costs, and control programs are often irregular or non-existent in significant portions of the country\cite{joshi2009vector}\cite{mondal2008vector}.
The overall driving forces of dengue prevalence in this country are: (a) lack of a broad intervention, (b) the lack of efficient human or electronic supervision, (c) inadequate data, and (d) lack of detailed investigations. These factors prompted us to investigate local newspaper data in order to  assess its potential as a supplementary surveillance system.
In the table ~\ref{tab:existingSurveillanceSystem}, we show some examples of similar systems proposed and implemented worldwide, while exploring their applicability in Bangladesh.

\section{Contribution}
As health related catastrophic events are published in local/regional languages with much care for the public awareness, the news source can be an alternative source for surveillance system in countries like Bangladesh. 
Since, the language barrier makes global monitoring platforms like HealthMap or Google Flu Trends incompatible \cite{schwind2014evaluation}, we explore the scope of using local newspapers and its potential to be a dengue surveillance system in Bangladesh.
Our main contributions in this paper include: 

\begin{itemize}

    \item \textit{Creation of a dataset of Dengue news}:  We gathered dengue-related news from 270 online Bengali newspapers and created a corpus of approximately 38,000 dengue disease and intervention-related news, in between 2017 to 2019. 
    We have cleaned and analyzed this corpus using human in the loop and well-known data mining technique for generating intervention reports on dengue for all districts in Bangladesh. 
    
    \item \textit{Identification of Intervention zones}: We identified the local regions that are likely to be deprived of proper governance of dengue intervention strategies by assessing the collected data from multiple demographic, environmental, and socio-economic perspectives. From the analysis of the data, we discovered that there is a significant disparity between the  authorities' intervention attempts.
    
    \item \textit{A demo website}: To help visualize the dataset, we created a website \textbf{(http://erdos.dsm.fordham.edu:3009)} which visualizes the relationship between official dengue records and news reports published in the digital domains. It supports data visualization for up to the sub-district level and can be used to explore past data for the selected area too. 
    
\end{itemize}

\vspace{-0.2cm}
\section{A Novel Dataset for Newspaper based Dengue Surveillance}
\subsection{Source}
To collect the daily archived dengue related news published in the newspaper, we used the dedicated web crawler of the Bangla Search Engine, Pipilika ~\cite{pipilika}. 
It uses the Apache Solr service to index and process queries. It has a collection of news data from 884 unique domains from 2017 to July of 2020. Details of the dataset volume is mentioned in the \textbf{Table ~\ref{tab:freqCD}}.
The final dataset is assorted into several granularity levels, based on different criteria. Spatially, we have subsets of the total dataset for each of the 64 districts, each of the 8 divisions, and each thana of Dhaka and Chittagong divisions. Temporally, these subsets are further subdivided for each month of the year for the range 2017-2019. 
The newspaper dataset was thoroughly pre-processed by removing special characters, removing stop words and URLs, and normalizing unicodes. To evaluate patterns related with our newspaper data for different parts of the country, we obtained daily cases of dengue patients from IEDCR, and a2i as well as socio economic data from UNDP. 

\begin{table}[!ht]
  \caption{Frequency of total crawled news vs total dengue related news}
  \label{tab:freqCD}
  \begin{tabular}{|c|c|c|c|}
    \hline
    Year&Total News Crawled&Dengue Related News&Percentage\\
    \hline
    2017 & 48780 & 950 & 0.19\% \\ 
    2018 & 1114701 & 2262 & 0.20\% \\ 
    2019 & 1754361 & 35796 & 2.04\% \\ 
  \hline
\end{tabular}
\end{table}

\subsection{Labeling through seed-guided LDA and human-in-the-loop: }

\begin{figure}[t]
    \centering
    \includegraphics[width=0.8\columnwidth]{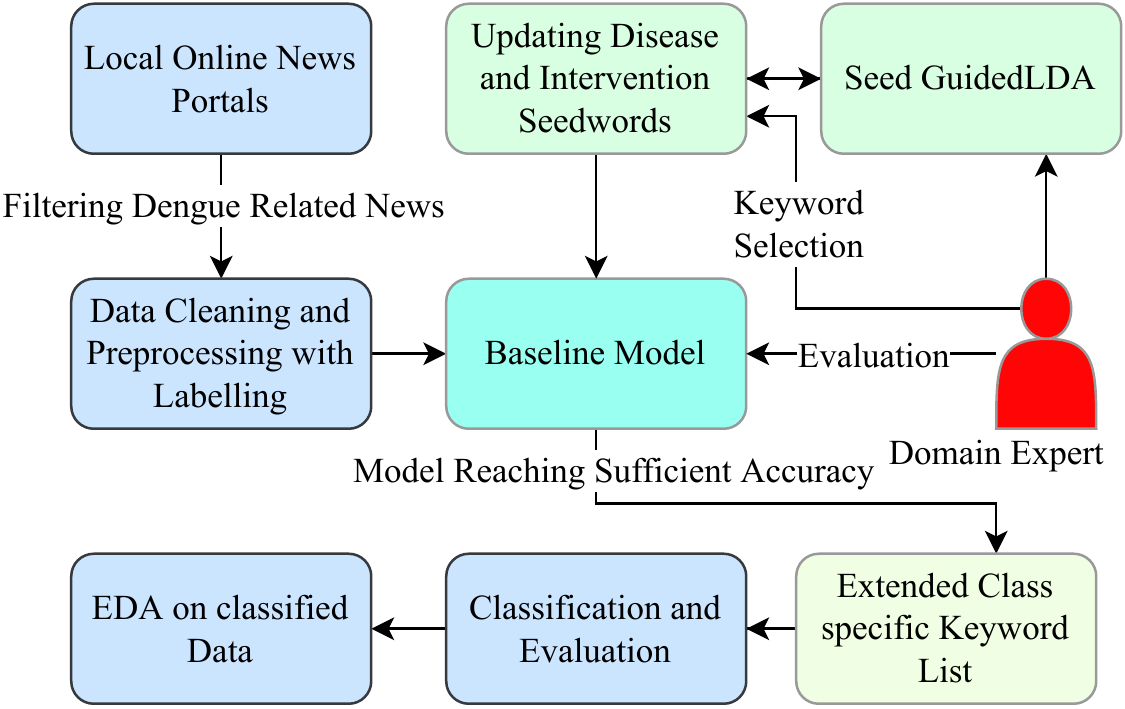}
    \caption{Workflow of dataset creation, curation and modeling for dengue disease and intervention news classification.} %
    \label{fig:workflow}
\end{figure}

To analyze our large corpus, and generate discernible outcomes for a surveillance system, we followed semi-automated human-in-the-loop (HITL) approach. We created a labeled dataset for the classification of disease and intervention related news. These labeled data are used for surveillance. 

\begin{figure}[!b]
    \vspace{-.2cm}
    \centering
    \includegraphics[width=0.6\columnwidth]{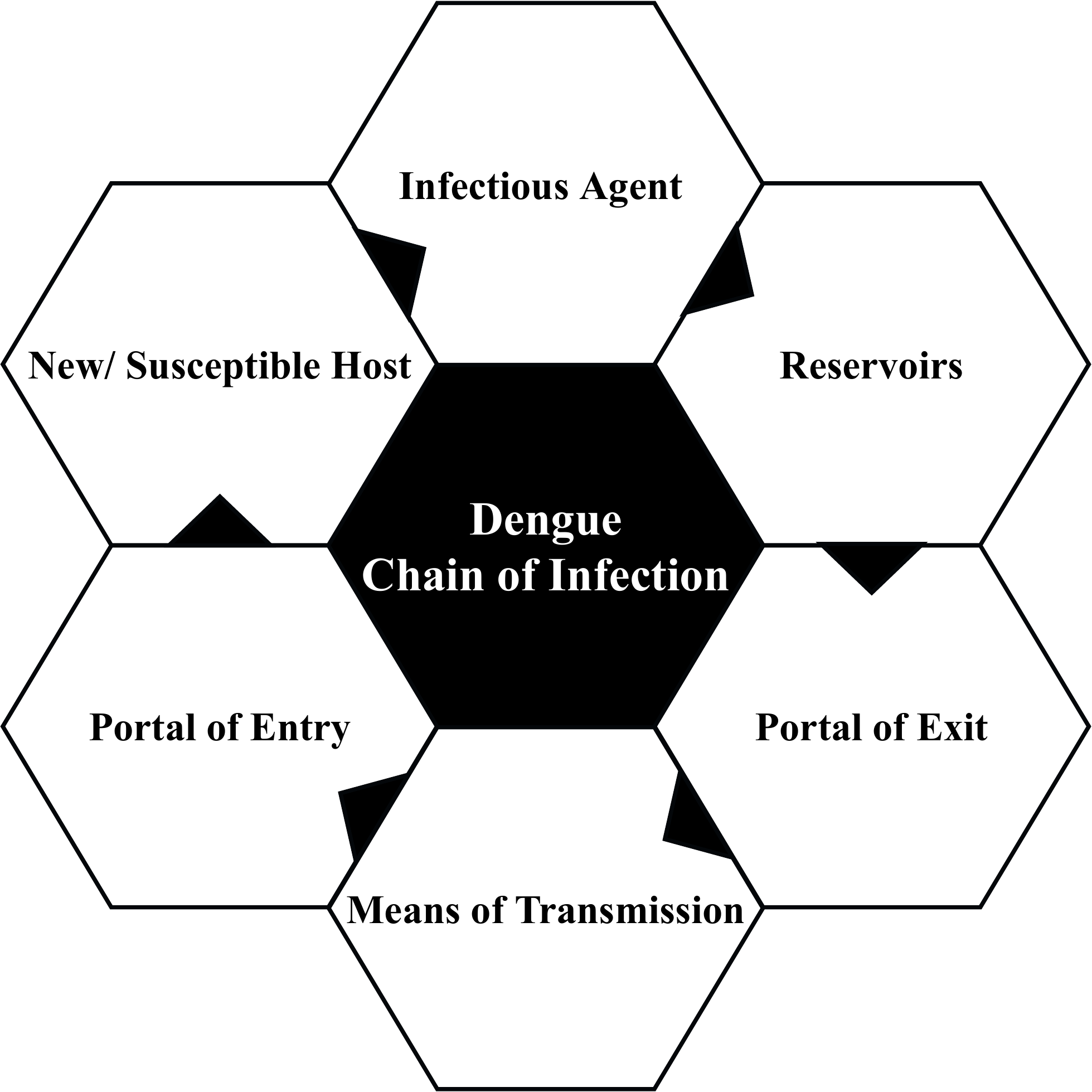}
    \vspace{-.1cm}
    \caption{Dengue Chain of Infection}
    \label{fig:coi}
\end{figure}


\subsubsection{\textbf{Seed Word Selection}} 

The spread of dengue infection can be summarized through a series of interactions among agents, hosts and environment, called the Chain of Infection (CoI) as presented in the \textbf{Figure ~\ref{fig:coi}}. 
The process starts with an an agent leaving its reservoir or host via a portal of exit. Then the agent transfers, by some method of transmission, and enters via a portal of entrance to infect a susceptible host. In the case of dengue, CoI relevant words could be \textit{dengue, DEN, drain, pond, ditch, blood, skin, mosquito, biting, human} etc. 
In collaboration with a public health domain expert, we created six sets of seed words that represent CoI from an initial list of 35 words. Those words were used as a prior for topic modeling and influenced the choice of words for each topic. We extended those sets of seed words by the application of the seed-guided LDA topic modeling over all the news data, followed by a filtering process according to the HITL technique \cite{jagarlamudi2012incorporating}, in multiple iterations (see \textbf{Figure \ref{fig:workflow}}).
At each iteration, the designated expert identified the latent topics within the dataset and selected most relevant words for each set of seed words. The iteration was terminated when the human evaluator could not extend the keyword sets anymore.

\subsubsection{\textbf{Justification of Two Classes for Labeling Dataset}}

To create a labeled dataset, we used the extended keyword sets. At the end of the HITL process, we had 60 additional terms along with the 35 seed words.
As the major decisions in public health measures depend on the infected cases count and the intervention outcomes, we decided to split our dataset into two classes: 1) \textit{Disease} representing general news regarding the infection and patient information; and 2) \textit{Intervention} representing reports offering details about occurrences, circumstances, or situations that highlight government intervention for cleaning or preventative measures. 
\textit{Disease} class comprises of the keywords mainly from infection agent, portal of entry/exit, and susceptible hosts. Whereas, \textit{Intervention} class comprise of the keywords mainly from reservoir, and means of transmission.

\subsubsection{\textbf{Creating Labeled Dataset}}

For labeling the dataset into two classes, we resorted to the HITL technique (\textbf{Figure ~\ref{fig:workflow}}) again. In this process, the designated human expert had been associated to label the dataset manually for the two defined classes of \textit{Disease}, and \textit{Intervention}. 
To expedite the HITL process for labeling the data, we created a baseline model using and ensemble of the Cosine and Jaccard similarity metric. 
For a given news data, we computed the Cosine similarity between the words in the news and the two keyword sets. If the similarity score was found to be over 50\% for any labels, the domain expert along with the two authors examined the news content. 
The label for the news was decided using majority inter annotator agreement. As some news have both set of keywords, human annotator involvement was necessary. We continued to create the dataset until we collected around 1,500 labeled records when the baseline accuracy using cosine similarity score reached $>80\%$.
Among them, 1,045 news reports belong to the \textit{Disease} class and 505 belong to the \textit{Intervention} class. 

\section{Training and Evaluation}
To classify our labeled corpus into the aforementioned two classes, first we limited the feature space of our text data only to the extended keyword set. These keywords correspond to the theme of each category definition.
While performing standard classifier training for our task, we used well-known and frequently used classifiers such as, Naive-Bayes classifier, KNN, and SVM with CountVectorizer as feature representation method. 
As demonstrated in the relevant literature studies, SVM classifiers were found to perform reasonably well for a smaller training and test dataset. We used a cross-validation ratio of 70\%-30\% and applied both binary and one-vs-rest mode. The overall system workflow can be found in the figure ~\ref{fig:workflow}. The performance of each classifier is summarized in \textbf{Table ~\ref{tab:tfidfscores}}.

\begin{table}[b]
\centering
  \caption{The performance of news classification model in identifying \textit{Intervention} and \textit{Disease} related news.}
  \label{tab:tfidfscores}
  \begin{tabular}{|p{5cm}|p{1.25cm}|p{1.2cm}|}
    \hline
    Method&Accuracy Score&F1 Score\\
    \hline
     Multinomial Naive-Bayes Classifier & 87.56\% & 87.65\%\\ \hline
     K Nearest Neighbours & 89.78\% & 90.60\%\\ \hline
     \textbf{SVM One-vs-Rest} & \textbf{91.56\%} & \textbf{91.45\%}\\ 
  \hline
\end{tabular}
\end{table}

\section{Results and Discussions}

\begin{figure}[t]
    \centering
    \includegraphics[width=0.8\columnwidth]{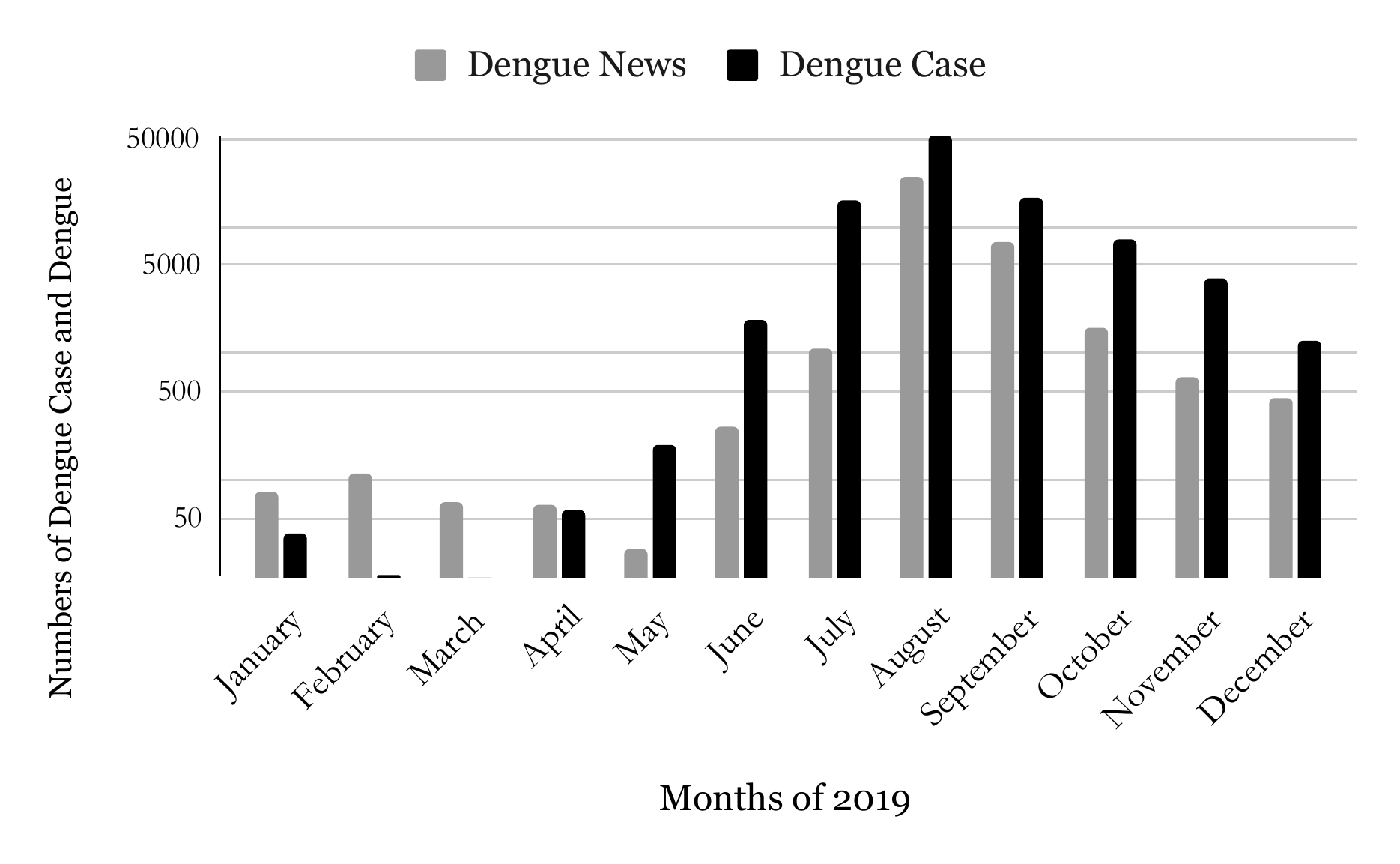}
    \caption{Countrywide Dengue cases showing a strong correlation between the frequency of officially recorded dengue cases and our collected dengue news reports.}
    \label{fig:country_case}
\end{figure}

\begin{figure*}[t]
    \centering
    \includegraphics[width=\textwidth]{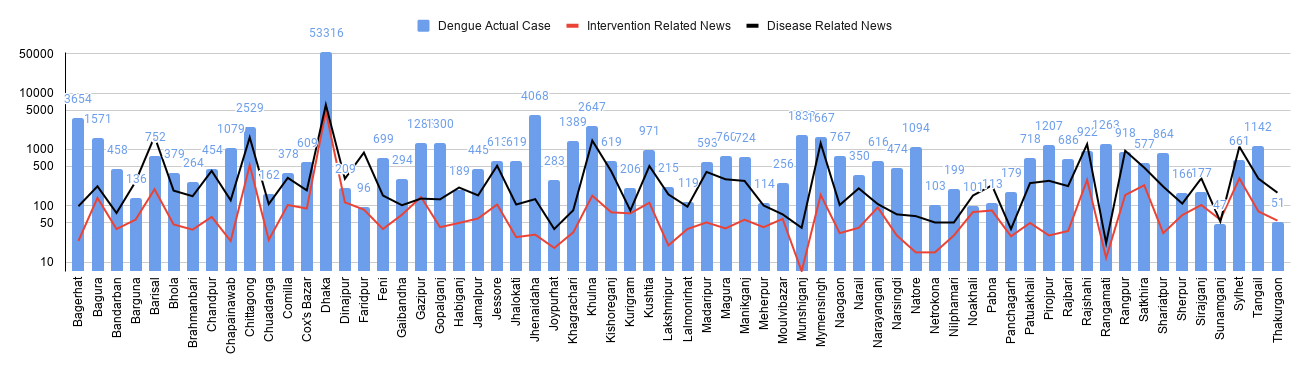}
    \caption{Analysis of recorded dengue cases and classified dengue related news in 2019, showing the discrepancies, where the intervention efforts are poorer in comparison with the number of actual cases, with blue color bars. It is also showing the strong attention that the divisional districts (Dhaka, Chittagong, Sylhet, Rajshahi, etc.) get in comparison with other districts.}
    \label{fig:district_cases}
\end{figure*}

On a large corpus of around 38,000 news stories on dengue, we applied the categorization methods outlined in the preceding section and divided them in two classes. We study several socio-economic and population parameters related to dengue incidence and their connection with published reports on dengue.
Our analysis shows that, given enough data, the number of dengue news strongly correlates with the number of dengue cases in real-time (see \textbf{Figure  \ref{fig:country_case}}). So this can act as an alternate medium to get the overall picture and intensity of dengue disease in different times. 

\begin{figure*}[!h]
    \centering
    \includegraphics[width=1.4\columnwidth]{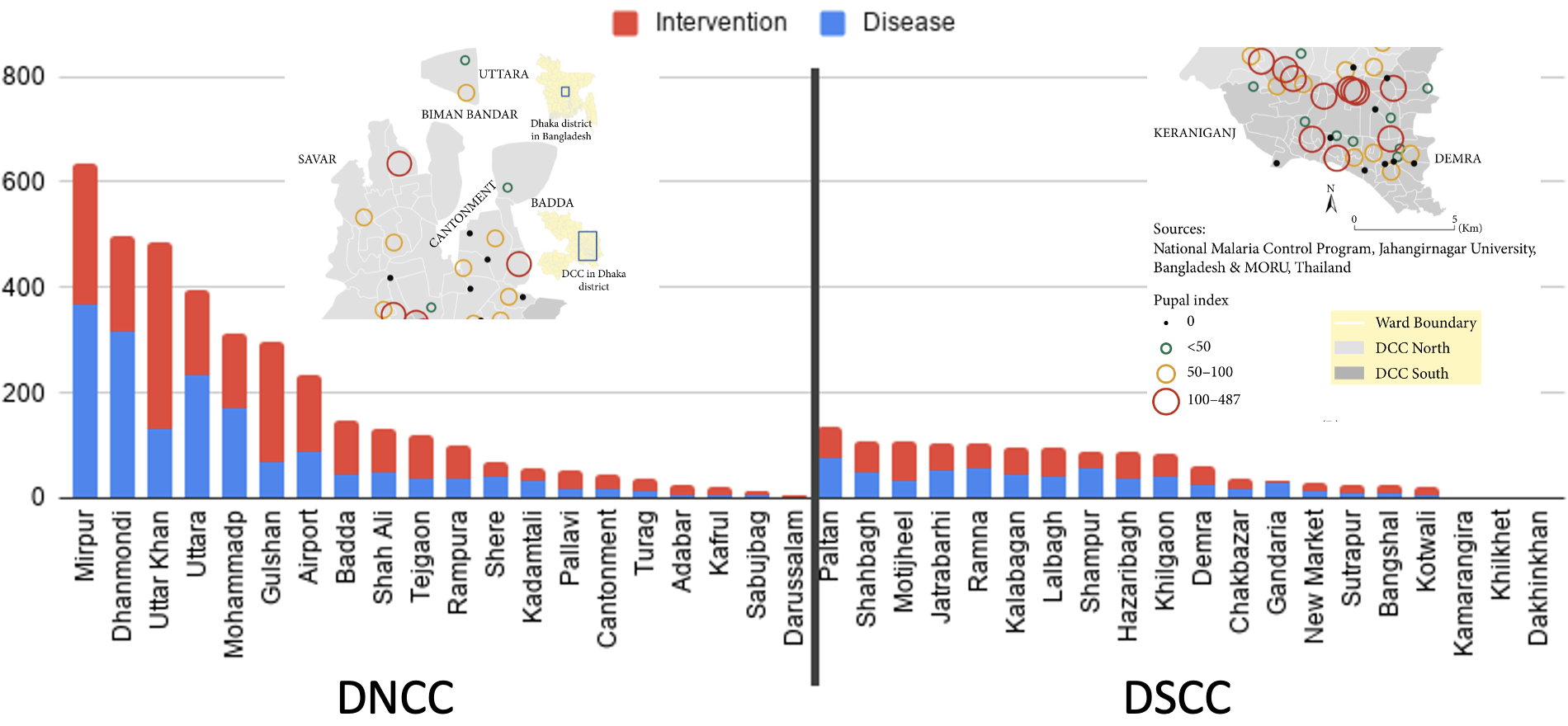}
     \caption{Dengue news analysis for the two city corporations in the Dhaka city shows that dengue disease and interventions news were reported more frequently in the DNCC than DSCC despite having more infection in DSCC \cite{mutsuddy2019dengue}}. 
    \label{fig:dncc_dscc}
\end{figure*}

We analyzed the dengue disease and intervention related news for each of the districts in Bangladesh for 2019 which is presented in \textbf{Figure \ref{fig:district_cases}}. It shows striking differences among the actual infected cases, disease report and intervention report among different districts. Since, divisional districts, such as Dhaka, Chittagong, Sylhet, Rajshahi, Khulna, Barishal are the financial hubs and densely populated, we see more dengue cases there. Newspapers reported the dengue disease and intervention news in similar proportion for those cities. But this plot also presents some districts, such as Bagerhat, Bandarban, Chapainwabganj, Feni, Jaypurhat, Lakshmipur, Munshiganj, Natore, Netrokona, Rangamati where despite having many dengue infected cases, we observe relatively lower number of intervention reports. Thus we suggest to revamp the intervention mechanism in those areas.

To analyze the disparity of actual dengue infected cases, and interventions reported in the newspapers at a more granular level, we analyzed the news of Dhaka city for individual areas that belong to two distinct city corporations, Dhaka North City Corporation (DNCC) and Dhaka South City Corporation (DSCC). DNCC and DSCC are different because of demographic, average income, education, and other socio-economic index according to UNDP. The human development index of DNCC is higher than that of DSCC.  In our previously published paper \cite{tasnim2021observing}, and also in \cite{mutsuddy2019dengue}, we showed that the distribution of dengue infected cases in the rich areas of DNCC is much lower than  that of densely populated lower income areas of DSCC. We found striking differences in the interventions related news reported for these two different parts. Although, news reports for rich areas are published frequently in general, the significant differences in disease and intervention news tells us that more interventions related to cleaning and maintaining to get rid of mosquito takes place in DNCC compared to DSCC (see \textbf{Figure \ref{fig:dncc_dscc}}).

\begin{figure*}[h]

    \centering
    \includegraphics[width=0.85\columnwidth]{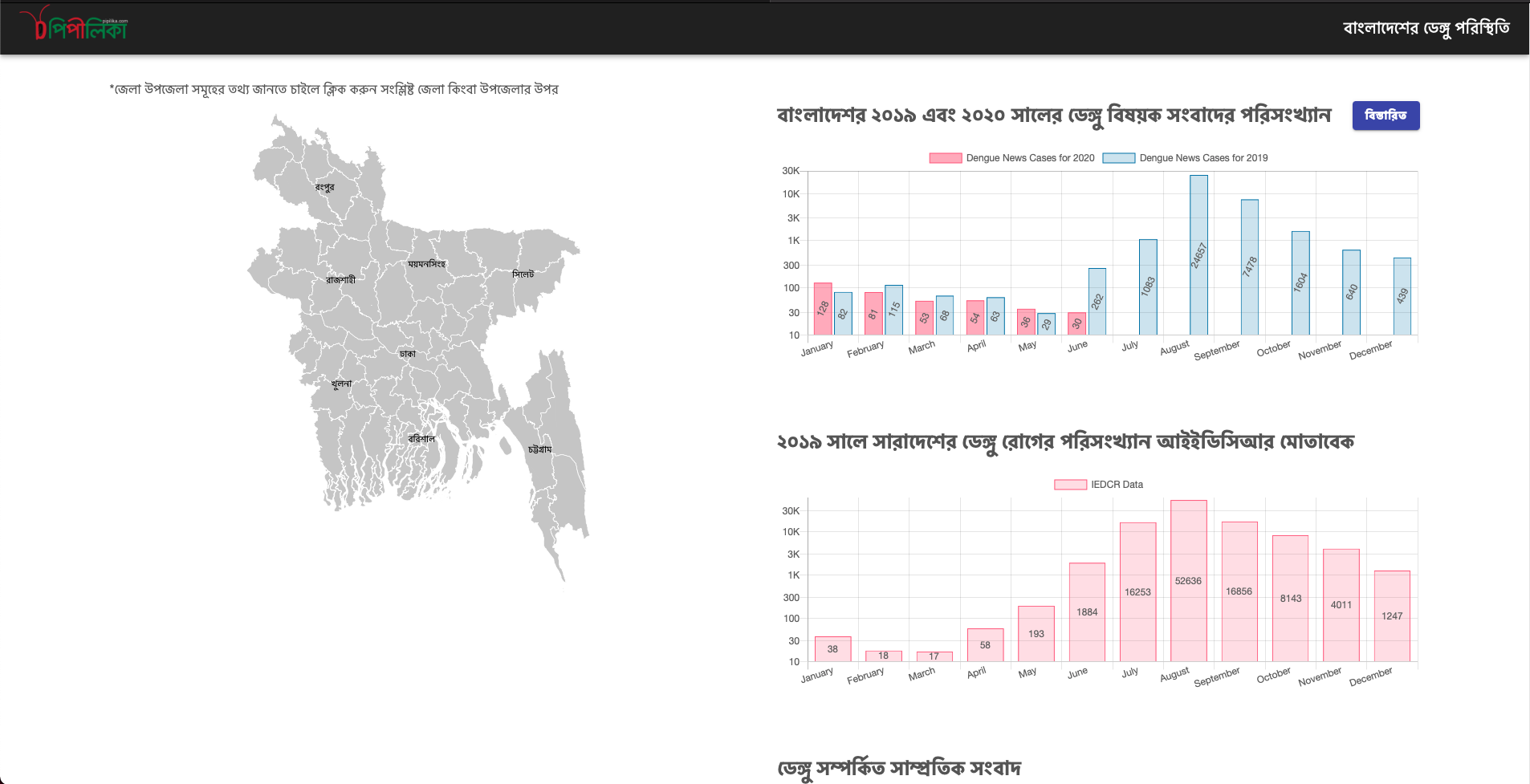}
    \includegraphics[width=0.85\columnwidth]{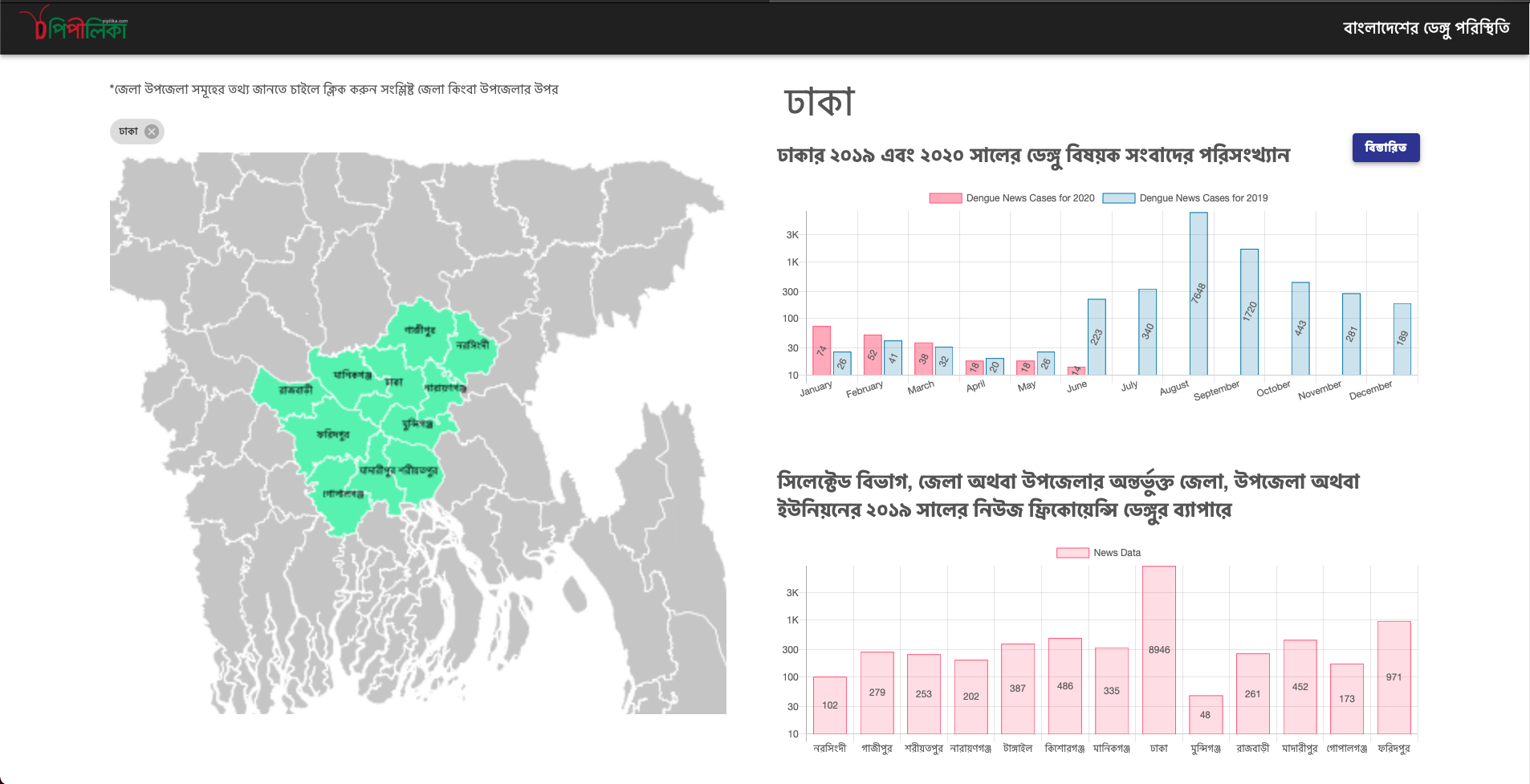}
    \includegraphics[width=0.85\columnwidth]{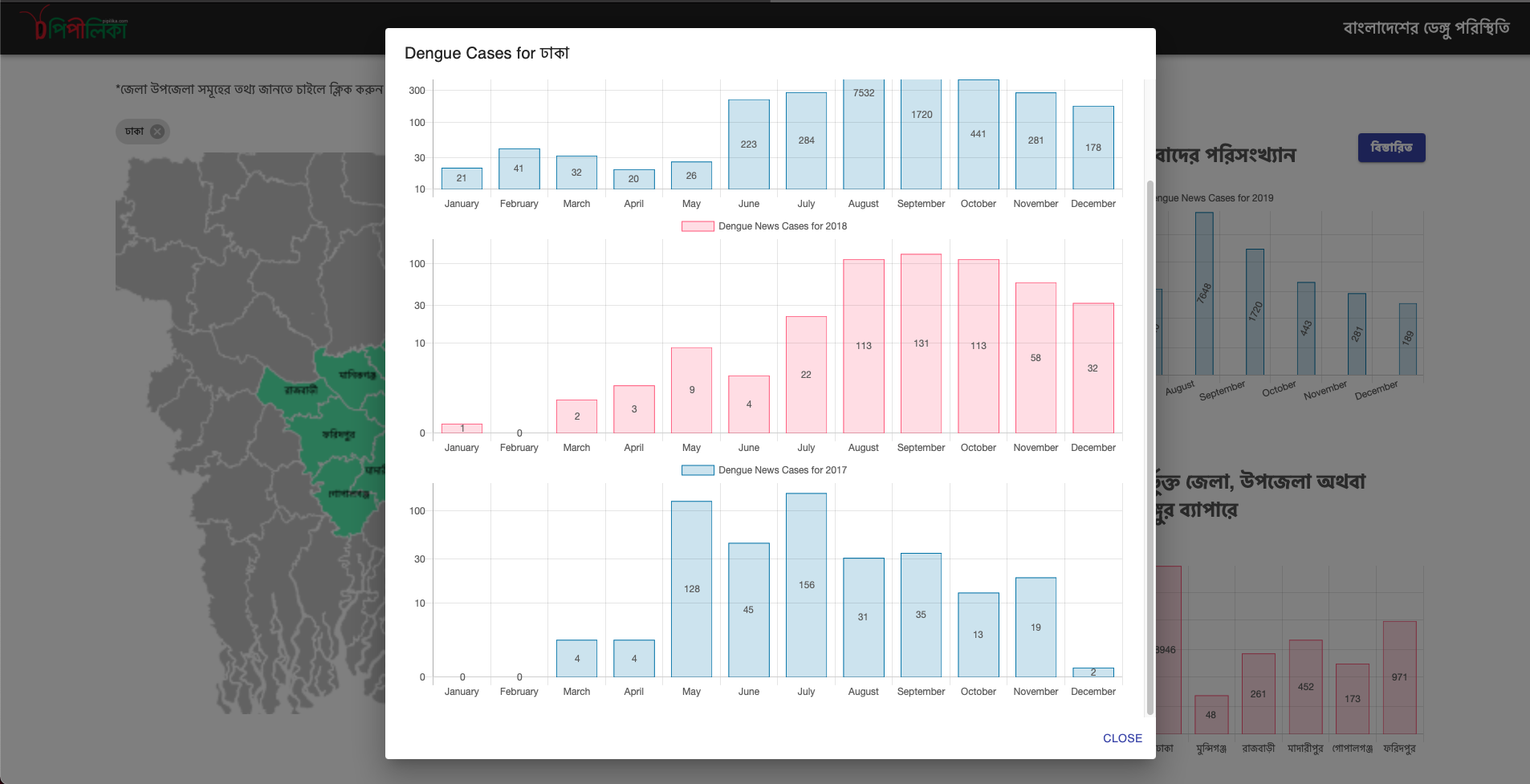}
    \caption{Preview of the visualization website. The first figure shows the overall data of the whole country and official dengue cases. The second figure shows the result of a selected district or sub-district level for further analysis. And finally, the third figure shows the preview of past data.}
    \label{fig:webpreview}
\end{figure*}

\section{Website Demo}

Some screenshots of the visualization website that we have created for this task are available in figure ~\ref{fig:webpreview}. The website was created primarily using \textit{ReactJS}. For plotting, we used \textit{d3JS}, and the API calls were completed through the \textit{axios} module.
The plots can be seen for division levels to more granular thana levels, for the years 2017 to 2019. Barcharts are used for easier comprehensions, while the actual values are visible on the mouse hover.

\section{Conclusion}

This paper focuses on creating a secondary surveillance system using published news on dengue. We outlined our methods and observation to create a dataset using HITL, described machine learning-based intervention and disease news detection, and finally generated observations that highly correlate with the dengue cases. Our analysis could identify the gaps in intervention mechanism, and can be used by the policymakers to amplify the public health support, in the deprived areas, to reduce the dengue infection rate.   
In this study, we had to limit our observation between 2017 -2019, and much of the demographic data were collected from third-party where in some cases the data was incomplete or insufficient. However, every year there are considerable increases in the amount of data (\textbf{Table ~\ref{tab:freqCD}}). We hope that our data-driven system will improve over  time, and become more reliable and meaningful as more and more data are collected. Thus, we conclude that online newspaper can be a suitable cost-effective supplementary to the official dengue surveillance system in a resource-constrained country like Bangladesh.

\bibliographystyle{ACM-Reference-Format}
\bibliography{0_manuscript}

\end{document}